\documentclass[aps,prl,superscriptaddress,twocolumn]{revtex4-1}
\usepackage[pdftex]{graphicx}
\usepackage[]{textcomp}
\usepackage[]{xspace}
\usepackage{dcolumn}
\usepackage{epstopdf}
\begin{document}

\title{Ultrasensitive measurement of micro cantilever displacement below the shot noise limit}

\author{R.C. Pooser$^{1*}$ \& B. J. Lawrie$^{1}$\\ $^1$Quantum Information Science Group, Oak Ridge National Laboratory, Oak Ridge, TN USA.\\$^*$pooserrc@ornl.gov}

\begin{abstract}
The displacement of micro-electro-mechanical-systems (MEMS) cantilevers is used to measure a broad variety of phenomena in devices ranging from force microscopes to biochemical sensors to thermal imaging systems. We demonstrate the first direct measurement of a MEMS cantilever displacement with a noise floor 4~dB below the shot noise limit (SNL) at an equivalent optical power. By combining multi-spatial-mode quantum light sources with a simple differential measurement, we show that sub-SNL MEMS displacement sensitivity is highly accessible compared to previous efforts that measured the displacement of macroscopic mirrors with very distinct spatial structures crafted with multiple optical parametric amplifiers and locking loops. These results support a new class of quantum MEMS sensor with an ultimate signal to noise ratio determined by quantum correlations, enabling ultra-trace sensing, imaging, and microscopy applications in which signals were previously obscured by shot noise.
\end{abstract}

\maketitle

\section{Introduction}

Optical beam displacement is a widely used measurement technique in micro and nano-mechanical sensors, imaging platforms, and force microscopes~\cite{rugar2004,arlett2011,lemieux2006}. 
The precision afforded by optical readout in these systems results in high signal to noise ratios (SNR), and in turn, inference of physical phenomena on the order of $\mathrm{fm/\sqrt[]{Hz}}$~\cite{fukuma2005,smith1995,putman1992}.
The combined noise in a MEMS cantilever displacement measurement serves to limit the sensitivity, so that a change in the frequency or amplitude of displacement brought on by, e.~g.,~the interaction with a single analyte molecule is impossible to discern if it falls within the noise statistics. While classical noise sources can be dramatically reduced with known technical approaches, the standard quantum limit (SQL) is a quantum mechanical noise limit stemming directly from the Heisenberg uncertainty principle as observed in two major noise sources, the back action noise and the shot noise level~\cite{Caves1981}:
\begin{equation}
\langle ( \Delta x )^2 \rangle_{SQL} = \langle ( \Delta x )^2 \rangle_{back} +    \langle ( \Delta x )^2 \rangle_{SNL} \label{eq:sql}
\end{equation} 
The back action noise arises from perturbations in the microcantilever position due to photon momentum-noise transfer.
In many cases, thermal noise, classical laser noise, and quantum mechanical back action can limit the minimum detectable displacement in microcantilevers, but it is possible to work in regimes that evade these noise sources~\cite{fukuma2005,krause2012high,hiebert2010,rutten2011high} so that the noise level of the coherent light field, called the photon shot noise limit (SNL), becomes the dominant source.

The SNL is determined by the Heisenberg uncertainty relation for minimum uncertainty states, and improvement in sensitivity beyond the SNL is impossible with classical optics \cite{Caves1981,jaekel1990}.  Light sources known as ``squeezed light'' displaying quantum-enhanced statistics below this limit for specific quadratures are available~\cite{Boyer2008,vahlbruch2008}, with applications in gravitational wave astronomy~\cite{schnabel2010} (using phase squeezed light) and bioimaging~\cite{taylor2013} (using amplitude squeezed light). Squeezed light has also been weakly coupled to microtoroidal resonators in order to realize a measurement of the resonator's displacement below the SNL~\cite{hoff2013} using phase squeezed light, but such an approach has not been generalized to microcantilevers or systems where strong coupling is possible. Optimal, sub-SNL beam displacement measurements in macroscopic systems have been realized by intricately crafting distinct optical spatial modes with multiple optical parametric amplifiers, beam combining cavities, and locking loops~\cite{treps2003,treps2002surpassing}, but the difficulty of such approaches combined with the difficulty in interfacing multi-spatial mode squeezed light with MEMS cantilevers has made a direct sub-SNL microcantilever displacement measurement impossible until now. Here we demonstrate beam deflection measurements with sensitivity below the SNL for typical AFM microcantilevers with a new technique that relies on a simple differential beam displacement measurement and two mode quantum correlations in the form of a two-mode squeezed state. This results in an accessible and stable approach to ultra-trace sensing, imaging, and microscopy with sub-SNL sensitivity. We demonstrate a noise reduction of 60\% below the SNL and state-of-the-art displacement sensitivity for a MEMS cantilever in the present experiment.

Importantly, our technique is enabled by accessing quantum correlations stored across multiple pairs of spatial modes, allowing for direct inference of microcantilever position based on optical beam deflection. The technique is markedly different from gravitometry, which uses phase squeezing and interferometry to detect mirror deflection, and other beam deflection techniques which craft designer single beam amplitude squeezed states in several spatial modes. In particular, using four wave mixing in rubidium vapor to construct the quantum-correlated fields allows us to craft distinct spatial modes by shaping the the pump field used in the nonlinear process. Using this technique we are able to conserve almost all of the initial spatial squeezing when performing position difference measurements on split detectors. The technique allows for a remarkably simple direct measurement on a single detector, while taking advantage of the noise reduction properties of differential measurements.

\section{MEMS noise sources}
Before reaching the SNL, the cantilever's various noise sources must be considered. Thermal motion decays inversely with oscillation frequency~\cite{fukuma2005}. In practical applications, displacement measurements performed with large spring constant microcantilevers at high frequencies not only reduce the thermal noise relative to other noise sources, but also enable high speed imaging and sensing applications~\cite{picco2007,rutten2011high}. There are many techniques in subsurface force microscopy~\cite{tetard2010new,shekhawat2005nanoscale}, time resolved force microscopy~\cite{sahin2007atomic}, stress imaging~\cite{tamayo2012imaging,wagner2011local,cuberes2000heterodyne}, ultrasonic force microscopy~\cite{yamanaka1994ultrasonic,dinelli1997ultrasound}, and generalized multi-frequency force microscopy~\cite{garcia2012emergence} that rely on off-resonant MEMS displacement measurements at frequencies where shot noise limited operation is achievable with commercial microcantilevers. Further, nearly shot noise limited operation close to the MEMS resonance frequency has also been demonstrated in custom optomechanical structures~\cite{krause2012high}. 

A growing number of manuscripts have demonstrated reduced quantum back action noise in cantilevers via laser cooling or other back action evading techniques~\cite{verlot2010,arcizet2006,teufel2009,anetsberger2010}, but the SNL provides a limit to sensitivity which cannot be improved upon with classical light. The variances due to back action noise and shot noise respectively are given by:
\begin{equation}
\langle ( \Delta x )^2 \rangle_{back} = \frac{8Ph\Delta f Q^2}{c\lambda k^2}; \hspace{0.1in}    \langle ( \Delta x )^2 \rangle_{SNL} = \frac{hc\lambda\Delta f}{8\pi^2P} \label{eq:noise}
\end{equation} 
where $x$ is the displacement, $P$ is the optical power, $\Delta f $ is the signal bandwidth, Q the microcantilever mechanical quality factor, $k$ is the spring constant, $h$ is Planck's constant, $c$ the speed of light, and $\lambda$ the optical wavelength \cite{smith1995}.

One approach to reducing the shot noise floor relative to the signal is to increase the optical power, which leads to higher SNR, but as Eq.~\ref{eq:noise} shows, doing so places limits on the sensitivity via increased back action noise. This leads to a conundrum in which the sensitivity cannot be arbitrarily improved by increasing the optical power far above mW levels unless back action evading techniques are utilized to reduce the SQL~\cite{verlot2010,arcizet2006,teufel2009,anetsberger2010} to the shot noise limit. In addition, higher optical power may saturate the optical detector, damage photosensitive ligands, or introduce excess laser noise that greatly exceeds both the SNL and the back action limits~\cite{fukuma2005}, limiting the viability of simply increasing optical power. On the other hand, applying quantum noise reduction to the read out light field when the SNL dominates the noise floor, as in the above-mentioned applications, results in displacement signals with higher SNR. In the present manuscript we demonstrate microcantilever displacement signals below the SNL for the first time.

\section{Experimental techniques}
\label{sec:exp}
\begin{figure}[b!]
\centering
\includegraphics[width=2.5in]{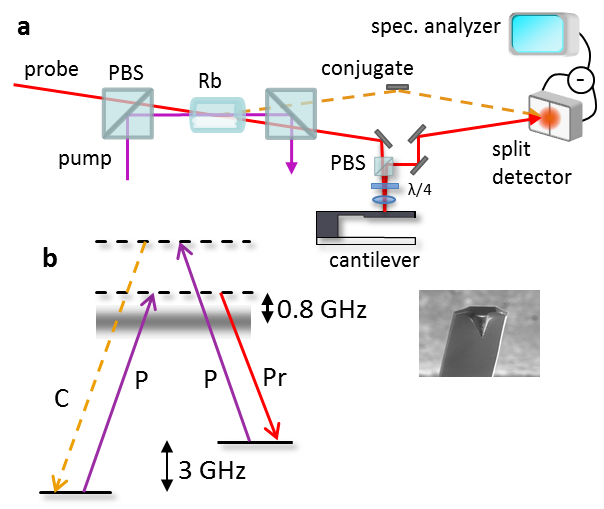} 
\caption{Differential beam position measurements with quantum correlated twin beams from four wave mixing. a) Squeezed differential beam displacement measurement using a split detector. The probe is sent to a polarizing beam splitter (PBS) where it passes through a quarter wave retarder before being focused onto the cantilever by a microscope objective. On the return pass the probe beam is separated at the PBS and sent to a split detector (or through an aperture), which subtracts the correlated noise between each half of each beam. b) The energy level diagram showing a double $\Lambda$ system at the D1 line (795~nm) in $^{85}$Rb. The presence of a weak probe (Pr) stimulates the coherent emission of a conjugate (C) photon for every emitted probe photon, while high nonlinear gain on the order of 5 allows for bright fields for both probe and conjugate.}
\label{fig:setup}
\end{figure}

Our detection technique relies on non-degenerate four wave mixing (4WM; see Fig.~\ref{fig:setup}) to produce twin beams with entangled spatial modes (sub-beam-size features) that each exhibit intensity difference quantum noise reduction. When incident on a spatially resolving detector such as a conventional split photodiode, the quantum correlated noise subtracts to yield a noise floor below the SNL for a differential measurement. As a result of the position-noise cancellation brought on by a differential measurement, no stabilization of laser frequency or pointing stability is required over periods of hours. 
In interferometric measurements with phase squeezing on the input port, the relative mirror displacements become anti-correlated, as the intensity difference noise is anti-squeezed after the first beam splitter~\cite{Caves1981}.
Our technique differs significantly in that the relative mirror displacements are not anti-correlated. Instead, a source of intensity difference noise reduction in multiple spatial modes occupies the input ports of the experiment, as shown in Fig.~\ref{fig:setup}, while the two beams are combined on a spatially-filtering detector which allows for a direct deflection measurement.  The squeezing angle is rotated exactly 90$^\circ$ relative to other techniques, resulting in highly sensitive microcantilever position measurements.

It has been shown that 4WM in Rb vapor can produce quantum noise reduction spanning multiple spatial modes~\cite{Boyer2008,Lawrie2013,qin2012}. The system has also lent itself well to integration with several sensing scenarios, including surface plasmon resonance sensors, quantum-enhanced image sorting, compressive quantum imaging, and nonlinear interferometry~\cite{lawrie2013prl,Lawrie2013,clark2012,hudelist2014quantum}. The ease of alignment afforded by the phase-insensitive amplifier configuration makes the generation of multi-spatial-mode quantum noise reduction accessible compared to implementations that use squeezed light in other sensors~\cite{taylor2013,hoff2013,treps2003}. In this experiment, a strong pump (150~mW) and weak probe (10 to 100~$\mu$W) mix at a slight angle ($0.3^\circ$) in a 12.7~mm long $^{85}$Rb vapor cell held at 130 $^\circ$C. The pump and probe are both focused into the center of the cell with 800 $\mu$m and 400 $\mu$m waists respectively.

During the 4WM process, for every probe photon emitted, a corresponding conjugate photon with opposite detuning is emitted, satisfying energy conservation. Likewise, the angle of emission is opposite that of the probe in order to conserve momentum. The Hamiltonian for the single spatial mode case is 
\begin{equation}
H = i\hbar\chi^{(3)} a_{1,k_1} a_{2,k_2} a^\dagger_{p,k_p}a^\dagger_{p,k_p}  + H.C., \label{H1}
\end{equation}
where $k_i$ denotes the field's spatial mode, $\chi^{(3)}$ is the nonlinear coefficient and $a_p$ is the pump field amplitude, which is assumed to be undepleted. The equations of motion are $\dot{\hat{a}}_1 =  \kappa \hat{a}_2^\dagger; \quad \dot{\hat{a}}_2 =  \kappa \hat{a}_1^\dagger$, which lead to the time varying operator solutions:
\begin{eqnarray}
    a_{1}(t) = a_1(0)\sqrt{G} - a_2(0)^\dagger\sqrt{G - 1}; \\ a_2(t)^\dagger = a_2(0)^\dagger \sqrt{G}-a_{1}(0)\sqrt{G - 1}
    \label{eq:gain2},
\end{eqnarray}
where $\kappa$ is the combined nonlinearity multiplied by the pump amplitude, which has been taken to be a classical number since it is undepleted and large in magnitude compared to the probe and conjugate, and where $\sqrt{G}=\cosh{\kappa t}$.
As a result, the probe and conjugate fields in identical spatial modes have quantum-correlated intensities. These quantum correlations manifest themselves as a lower noise floor in a measurement of the intensity difference between the beams (normalized to the SNL, in the case of no losses):
\begin{equation}
\langle\Delta (N_-)^2\rangle=\frac{1}{(2G-1)}, \label{eq:squeezing}
\end{equation}
where $N_-$ is the photon number difference operator.

In addition, macropixels of each beam known as coherence areas~\cite{Jedrkiewicz2004} are correlated pairwise across the beams. In the limit that each coherence area could be described by a single spatial mode,  the Hamiltonian would consist of multiple concurrent nonlinearities:
\begin{equation}
H = i\hbar\chi^{(3)} \sum a_{i,k_i} a_{j,k_j} a^\dagger_{p,k_p} a^\dagger_{p,k_p} + H.C. \label{H2}
\end{equation}
It has been shown that Eq.~\ref{H2} leads to quantum noise reduction for multiple modes in the time domain~\cite{Bennink2002} and the spatial domain~\cite{Marino_PM_PRA2013,Pooser2014}. The coherence areas are effectively independent and can be treated as spatial modes in Eq.~\ref{H2} if the coherences in the far field tend to zero; that is, if $\langle a(k_i)^\dagger a(k_j)^\dagger a(k_i) a(k_j) \rangle \rightarrow0$. If the coherence areas contained within the beams do not interfere with one another in the detection plane, then this condition is effectively fulfilled. 

\begin{figure}
\centering
\includegraphics[width=1.75in]{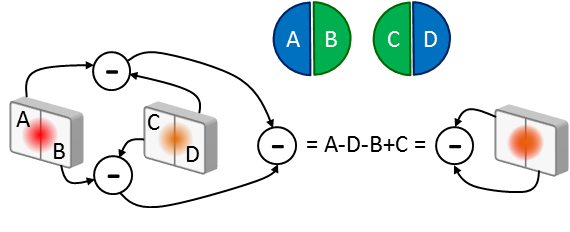} 
\caption{Relative beam position measurement of the probe and conjugate fields. The shaded semi circles represent correlated sub parts of each beam. When accessed with a split detector, the quantum-correlations in these modes reduce the total noise floor. Because the probe and conjugate are mirror images of one another about the propagation axis, a single split detector (right hand side) can be used to access the position correlations that a relative position measurement would reveal (left hand side).}
\label{fig:diffdet}
\end{figure}

If each pair were isolated and the intensity difference measured, the quantum noise reduction would approach that of Eq.~\ref{eq:squeezing}. A split photodiode placed in each beam's image plane would perform this measurement when properly aligned. In particular, the probe and conjugate beams illustrated in Fig.~\ref{fig:setup} demonstrate 4.5 dB of squeezing when isolated on individual photodiodes, but maintain 4 dB of squeezing when evenly distributed on the split photodiode.  This indicates that - to a good approximation - the coherence areas in the probe and conjugate beams are well isolated on each side of the split detector.  The ideal model is shown in Fig.~\ref{fig:diffdet}. The conjugate spatial modes contained in region \textit{A} are quantum correlated with the probe spatial modes in region \textit{D}, and the conjugate spatial modes in \textit{B} are quantum correlated with the probe spatial modes in \textit{C}. While neither beam contains the perfect split noise mode ideal for beam displacement~\cite{pinel2012,treps2003}, the beams contain sufficient spatial information to show 60\% reduced uncertainty in the relative displacement compared to a measurement using classical light. We also note that the differential position measurement is compatible with homodyne detection as a drop in replacement for split detectors, and that the sensitivity achievable in the split detector approach is equivalent to that possible with interferometric detection~\cite{Barnett2003}. In comparison to weak measurement techniques~\cite{hosten2008,dixon2009}, our approach is deterministic and does not require post-processing. 

Finally, we note that perfect alignment is not always possible. The coherence areas in reality are not as neatly distributed as in Fig.~\ref{fig:diffdet}. However, the worst outcome in this scenario is that the spatial modes that are distributed over the detector halves contribute only fractions of shot noise units to the measurement noise floor, since spatial modes that do not sum to the flipped TEM mode contribute units of shot noise to the difference measurement~\cite{treps_multipixel,Barnett2003}. In this case, the noise, normalized to the shot noise level, is given by
\begin{equation}
\langle \Delta N^2 \rangle =\frac{1}{P_0}\left[\frac{P_{s}n_s}{( 2G-1) }+\left(\sum_{i=1}^{M} \frac{P_{i}(x)n^\prime_{i}(x)}{\eta_{d}(2G-1)} \right)\right],
 \label{eq:x2}
\end{equation}
where $P_{s}/P_{0}$ is the fraction of optical power in coherence areas isolated in one mode of the detector,  $P_i(x)$ is the power partially incident on a detector mode from the $i^{th}$ optical mode such that $\sum_{i=1}^{M} P_{i}(x) =P_{0}-P_{s}$, $n_s=2G-1+2\eta_d-2G\eta_d$ is the twin beam noise for single spatial modes, $\eta_{d}$ is the combined detector efficiency, $M$ is the total number of modes split across the detector, $n^\prime_{i}(x) = \eta_i(x)(2G-1+2\eta_i(x)-2G\eta_i(x))$ is the noise of the $i^{th}$ mode, where $\eta_i(x)$ depends on misalignment, and it has been assumed that each amplifier mode has equal input amplitude and equal gain. Note that the position dependence in $P_i(x)$ and $\eta_i(x)$ depends on the specific overlap of each spatial mode with each detector half. The most important feature of Eq.~\ref{eq:x2} is that a multimode beam always results in a lower noise floor than a single mode beam (for a single mode beam split in half, the noise doubles vs the ideal multimode case, for instance). In the limit that the sizes of the coherence areas are much smaller than the detector halves ($P_{s}$ approaches $P_0$), the quantum noise reduction approaches the ideal case.
\begin{figure*}[t]
\centering
\includegraphics[width=6.5in]{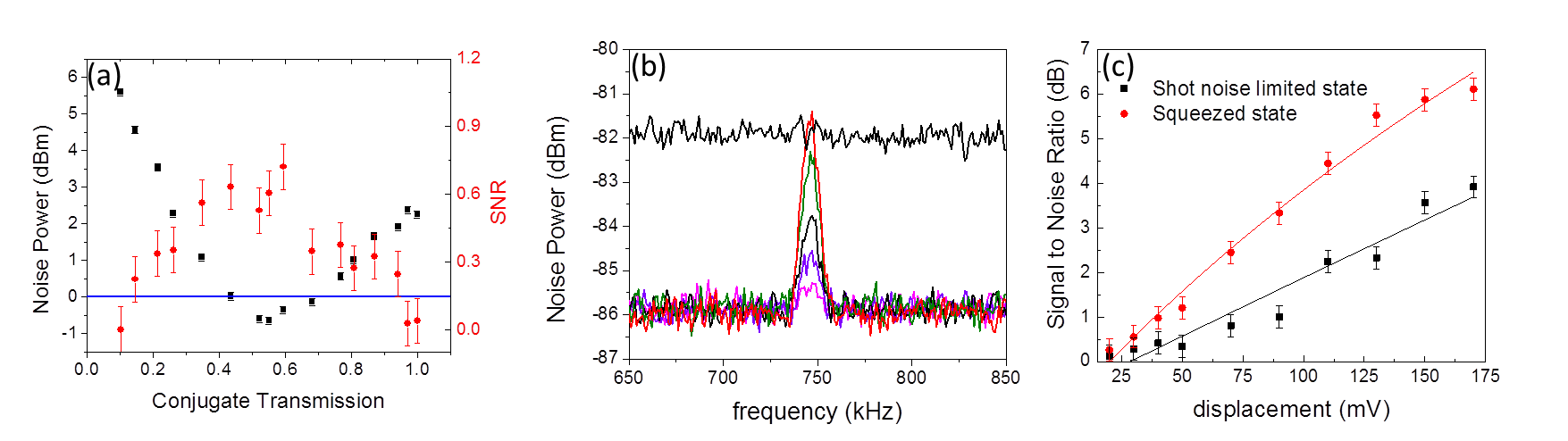}
\caption{Sub shot-noise microcantilever deflection detection. (a) Sub shot noise modulation for a microcantilever with an aperture in the beam path,  SNL-normalized noise-power (left axis, $\pm$0.1 dB error bars, standard deviation of 1000 averages) and SNR (right axis) as a function of conjugate transmission through an aperture with a constant probe transmission of 50\% ($\pm$0.1 dB standard deviation of 1000 averages), (b) Spectrum analyzer traces of raw displacement signals on a split detector centered at 745~kHz for various displacement amplitudes, from 30~mV to 120~mV. The noise floor shows a broadband squeezing level of 4.0$\pm0.1$~dB relative to the shot noise level (black line, electronics noise subtracted). The spectrum analyzer settings were: resolution bandwidth: 10~kHz; video bandwidth: 100~Hz; sweep time: 2~s; 20 averages, and (c) Signal to noise ratio (with error bars of $\pm$~0.2~dB, statistical uncertainty) on a split detector vs increasing displacement amplitude at 745~kHz for a squeezing level of 3.0$\pm$0.1~dB. The squeezed SNR increases more rapidly than the coherent case to a separation of 3~dB in the limit of large displacement.}\label{fig:MC_sqz}
\end{figure*}

\section{Microcantilever displacement measurements}
The measurements reported in this paper were performed on a gold-coated microcantilever with a fundamental resonance at 13 kHz and a force constant of 0.2 N/m (BS-contGD from NanoAndMore) mounted in a Bruker AFM mount with a piezo-driven actuator connected to a function generator. The piezo-cantilever system has a combined resonance due to piezo-loading, which modifies the frequencies at which the device will resonate. One such resonance was found at 745kHz, where the displacement measurements presented here were performed. This particular resonance showed a quality factor of 124.
For a typical optical power of 130 $\mu$W, the cantilever was shot noise limited at ambient temperature and pressure for oscillation frequencies above 400~kHz, with an SNL of 3.9~$\mathrm{fm/\sqrt[]{Hz}}$ and a quantum back action limit of 33~$\mathrm{am/\sqrt[]{Hz}}$. The SNL exceeded the back action noise for optical power less than 15 mW. Our source would still generate squeezed states above 15~mW with sufficient input probe power, but back action noise would determine the noise floor. Likewise, for measurements performed at low temperature and pressure where thermal noise is dramatically reduced even at low frequencies, our approach can be applied using low frequency squeezed states~\cite{Mccormick2008low,liu2011} to further reduce the noise floor near a fundamental resonance.  

Two detector configurations were used to detect a position modulation generated by driving the microcantilever with a piezo-controlled tapping mode AFM mount at 745~kHz. First, razor blades were used as apertures in the beam paths in order to simulate a split detector and to aid in aligning the multi-spatial-mode contribution to the displacement measurement. As illustrated in Fig.~3(a), this configuration allowed for the direct measurement of microcantilever beam displacement with sensitivity beyond the shot noise level.  When a razor blade was used to attenuate 50\% of the probe beam, corresponding to the highest SNR for classical light readout with apertured detectors, and a second razor blade was rastered across the conjugate, it was evident that maximum signal to noise ratio was coincident with maximum squeezing. This provides clear evidence that squeezed states exceed the sensitivity of coherent states, even when 50\% attenuation is artificially introduced. On the other hand, when the spatial mode distribution of the probe was changed in its image plane relative to the conjugate by introducing spherical aberrations at the cantilever microscope objective, there was no enhancement in SNR over the classical case, demonstrating that the quantum correlated spatial modes between the two beams were responsible for the reduced noise floor in the deflection measurement. 

We used the razor blades as an aid in alignment on a position sensitive detector by ensuring that both the probe and conjugate image planes were coincident on the detector and that no spherical aberrations were present. While this method proves that sub-SNL displacement measurements are possible, the gains are reduced due to absorption of 50\% of the probe, and while apertured detectors are commonly used in displacement measurements, the artificial introduction of loss means that they can not provide maximum signal to noise. Indeed, the minimum detectable displacement also depends on the losses~\cite{fukuma2005}. The SNR and measured noise level outlined in Fig.~3(a) correspond to a maximum increase in SNR of 0.7~dB versus the classical case, in line with what has been achieved with optimal measurements in other MEMS systems.  Below, we demonstrate a dramatic improvement on this enhancement by utilizing high quantum efficiency split photodiodes in place of the apertured photodiodes.

The balanced split detector illustrated in Fig.~\ref{fig:diffdet} yields the SNR associated with traditional split detector schemes~\cite{fukuma2005} and the common mode rejection associated with traditional balanced detection schemes. The individual diodes in each split detector have 96\% quantum efficiency, and utilizing a single split diode to measure relative probe and conjugate beam position ensures that the diodes are ``matched'' (having quantum efficiency and terminal capacitance as close as possible to one another). When using separate split diodes for each beam, a unique possibility in this experiment, the electronic gain and frequency response of each can be tuned to account for any mismatch with the added advantage that the probe and conjugate fields can be space-like separated by a large distance.

The data presented in Fig.~3(b) and (c) were acquired with probe and reference beam combined on a single split detector, and the microcantilever was again modulated at 745 kHz with an amplitude of 20-170~mV.  Figure 3(b) shows data for amplitudes of 30-120~mV with a noise floor 4.0$\pm0.1$~dB lower than would be possible with classical readout light at the same optical power. In the limit of large displacements, the SNR is determined by the amount of quantum noise reduction available in the readout light, meaning that we directly observed an increase in SNR by 4~dB. For the 10kHz resolution bandwidth used for these measurements, this results in a reduction of the minimum resolvable cantilever displacement from 392$\pm1.2$~fm to 156$\pm1.2$~fm (1.56 fm$/\sqrt{Hz}$) - enabling the measurement of displacements previously obscured by shot noise.  In Fig.~3(c), the measured signal to noise ratio is plotted as a function of microcantilever modulation, illustrating the same effect for a squeezing level of 3.0$\pm0.1$~dB. For a given SNR, the squeezed measurement can always discern smaller displacements than coherent light. To the extent that squeezing exceeds the single-beam SNL for an equivalent power, as we have demonstrated here, a readout strategy that uses squeezed light will always yield better results than an equivalent readout strategy using coherent light alone.

\begin{figure*}[ht!]
\centering
\includegraphics[width=7in]{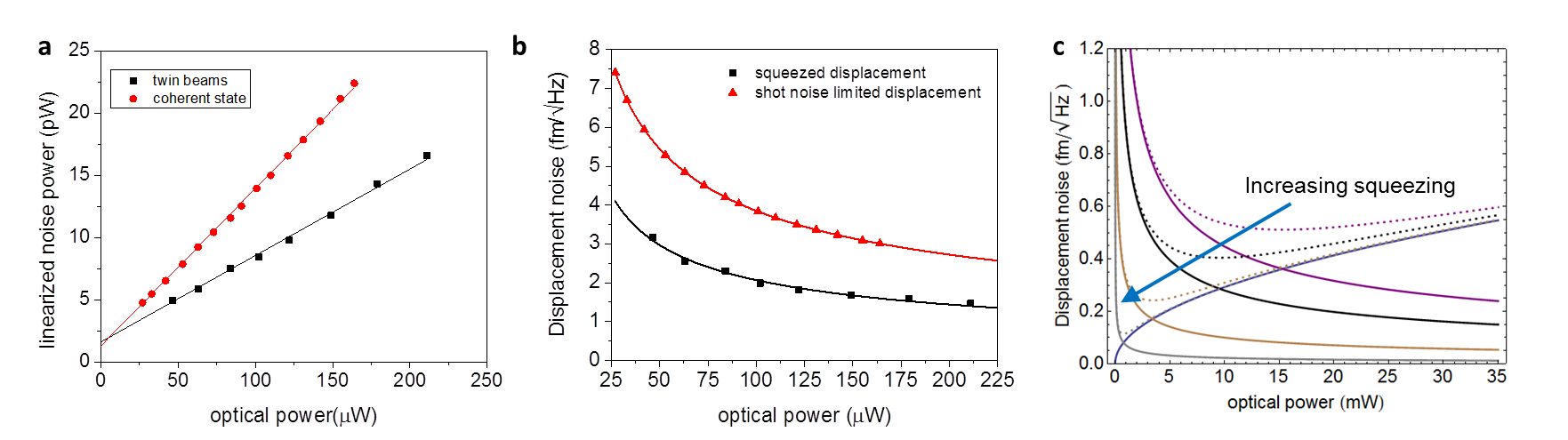} 
\caption{a) Noise floor as a function of optical power for split detector measurements using coherent states (circles) and squeezed states (squares). b) The minimum inferred displacement measurable, calculated using the SNR and measured noise floor. The curves are fits to the data from the theoretical calculations in (c), following the methods in~\cite{fukuma2005},~\cite{Fabre_quantum_limits}, and~\cite{Fujihira_cal}, which shows the optical noise floor, back action (blue), and SQL (dashed lines) as a function of optical power for various levels of squeezing. The purple curve is the shot noise (no squeezing), while the black, brown, and gray correspond to 4, 13, and 26dB of squeezing respectively.}
\label{fig:disppower}
\end{figure*}

Figure \ref{fig:disppower}a illustrates the noise floor for the position difference of two coherent states as a function of power along with the same measurement using two quantum-correlated beams showing 2.8~dB of quantum noise reduction. This plot also serves to demonstrate the shot-noise-limited detector calibration. The minimum discernible displacement is calculated from the SNR and noise data and plotted in Fig.~\ref{fig:disppower}b. The inferred displacements in this plot correspond to noise floors between 2.5 and 2.8~dB below the classical limit.

Figure~\ref{fig:disppower}c shows the theoretical noise floors, in the limit that additional back-action is negligible, for various levels of quantum noise reduction, including for 4dB, the maximum level observed in our experiment. Quantum noise reduction reduces the optical noise floor below the shot noise level by a factor exponential in the squeezing parameter, $e^{-r}$~\cite{Barnett2003}. In the case of infinite squeezing, the optical noise is insignificant compared to the back action limit (the 26dB case effectively illustrates this). The 4dB case, which corresponds to our experiment, results in more modest gains, but nonetheless falls below the shot noise, moving the back action limit down to 10~mW for this system. We make two other remarks concerning Fig.~\ref{fig:disppower}c. First, we plotted the theoretical noise floor for various squeezing levels assuming that detection of quantum noise reduction did not increase the back action noise. It has been pointed out that in certain detectors, such as interferometers which use squeezed vacuum on the input, the crossing point between optical noise would move to lower powers as in Fig.~\ref{fig:disppower}c, but the back action would also increase an amount commensurate with the squeezing~\cite{Caves1981}. In this case the SQL curves would increase more rapidly with power than shown here. On the other hand, it has been pointed out that squeezing in the proper quadrature does not induce additional back action in the variable of interest~\cite{jaekel1990}. Second, it is clear that, as expected, quantum noise reduction brings benefits in the region where the SNL dominates, to the left of the back action crossing point. This means that low-power sources of quantum noise reduction can be of great use, especially as characteristic sizes shrink to the nanoscale.

Finally, it is notable that some drift in the DC displacement of both the probe and conjugate beams is present due to air currents within the Rb cell. However, due to momentum conservation, the probe drift is canceled by an equivalent drift in the conjugate beam. Any residual differential drift is negligible at the 745 kHz displacement frequency under study. The pointing correlations are reliable enough to ensure a stable measurement over the course of several hours with no stabilization of any sort in the experiment. 

\section{Conclusion}
To conclude, the use of multi-spatial-mode squeezed states provides the first direct measurement of microcantilever beam deflection below the SNL. The results are applicable to any cantilever device operating with minimal back action noise, including optically transduced nano-electro-mechanical-systems (NEMS), which have recently been shown to be shot-noise-limited~\cite{hiebert2010} at frequencies of a few MHz. The 4WM process has approximately 50~MHz of bandwidth within which to demonstrate sub-SNL displacement for NEMS and MEMS structures. The experimental configuration is simple compared to most quantum sensing applications which require multiple optical cavities, cavity length locks, and phase locks. Here, no stabilization is required; instead the twin beam correlations cancel any excess position-noise. Further, interference measurements, while possible in this configuration, are not necessary to achieve high SNR displacements. In addition, the 4WM process supports arbitrary probe spatial profiles, allowing for additional improvement in sensitivity via the choice of appropriate beam profiles by placing a spatial light modulator before the vapor cell~\cite{Lawrie2013}.

\section*{Funding Information}
This work was performed at Oak Ridge National Laboratory, operated by UT-Battelle for the U.S. Department of energy under contract no. DE-AC05-00OR22725. B.J.L was supported in part by a fellowship from the IC postdoctoral research program. R.C.P. acknowledges support from the Oak Ridge Laboratory Directed Research and Development program.

\end{document}